\documentclass[times,10pt,twocolumn]{article}

\usepackage{latex8}
\usepackage{times}
\usepackage[dvips]{graphicx}
\usepackage{url}
\usepackage{float}
\usepackage{amsmath}
\usepackage{amssymb}
\usepackage{centernot}
\usepackage{hyperref}
\usepackage{mathrsfs}
\usepackage[dvipsnames, svgnames]{xcolor}
\usepackage{colortbl}
\usepackage{cite}
\usepackage{longtable}
\usepackage{layout}
\usepackage{authblk}
\usepackage{etoolbox} 
\usepackage[utf8]{inputenc}
\usepackage{cleveref}
\usepackage{listings}
\usepackage[utf8]{inputenc}
\usepackage{tabularx}
\definecolor{skyblue}{HTML}{87cefa}

\begin{document}

% Applying Model-based Statistical Testing for Cloud-based APIs
% Rigorous Software testing for Scientific Computing
% Lin: Applying Rigorous Testing Methodologies to RESTful APIs for Hydrological Modeling
% Applying Rigorous Testing Metholodologies to a 

% Song: Automated Statistical Testing and Certification of a Reliable Scientific Model Coupling Server

% Applying Rigorous Testing Metholodologies to Model Coupling Servers and RESTful APIs
% Statistical Testing Automation for the Coupling of Scientific Models 
% 

\title{Automated Statistical Testing and Certification of a Reliable \\Model-Coupling Server for Scientific Computing}

\author{%
  \begin{center} % Center the entire author block
    % Line 1: Author names
    Seth Wolfgang$^{1}$, Lan Lin$^{2}$, Fengguang Song$^{1}$ \\% Add a small space after the names

    % Line 2: Affiliation 1 (shared by Seth Wolfgang and Fengguang Song) and its location
    \normalsize \textit{$^{1}$Luddy School of Informatics, Computing, and Engineering, Indiana University, Bloomington, IN, USA}\\ % Add a tiny space after affiliation

    % Line 3: Emails for Affiliation 1
    \small{\{seawolfg, fgsong\}@iu.edu}\\
    
    % Line 4: Affiliation 2 (for Lan Lin) and its location
    \normalsize \textit{$^{2}$Department of Computer Science, Ball State University, Muncie, IN, USA}\\ % Add a tiny space after affiliation

    % Line 5: Email for Affiliation 2
    \small{llin4@bsu.edu} 
  \end{center}
}

\maketitle
\thispagestyle{empty}

\begin{abstract}
Sequence-based specification and usage-driven statistical testing are designed for rigorous and cost-effective software development, offering a semi-formal approach to assessing the behavior of complex systems and interactions between various components.
While this approach has been successfully applied to a number of domains ranging from medical devices to scientific instrumentation, it is particularly valuable for scientific computing applications in which comprehensive tests are needed to prevent flawed results or conclusions.
% Sentence two which is more specific to the problem
As scientific discovery becomes increasingly more complex, domain scientists couple multiple scientific computing models or simulations to solve intricate multiphysics and multiscale problems. 
These model-coupling applications use a hardwired coupling program or a flexible web service to link and combine different models.
In this paper, we focus on the quality assurance of the more elastic web service by automatically generating, executing, and evaluating 5,204 test cases via a combination of rigorous specification and testing methods.
% We certify the model-coupling server controller with a projected reliability estimate, and verify its correct functionality through an automatically generated-executed-evaluated test suite. % to guarantee proper data handling 
% map to the epilogue - broader picture 
The application of statistical testing exposes problems ignored by pre-written unit tests and highlights areas in the code where failures might occur. 
% Additionally, a quantitative measure of the server controller’s reliability is derived, offering a statistic to support a claim of its robustness and dependability.
We certify the model-coupling server controller with a derived reliability statistic, offering a quantitative measure to support a claim of its robustness. 
\end{abstract}

\section{Introduction}
% \note{Introduce the acronym for DES} \\
% \note{sign post here.}
% \note{Contributions: \\
% Mealy machine building, \\
% Auto testing implementation, \\
% Application of rigorous methods to scientific computing, \\
% minor one: find bugs and provide certification.\\
% minor one: Embarrassingly parallel testing}

% (extendable) automated testing, finding faults, applying method to this domain
% \todo{Brief paragraph to talk about E3SM}
% \todo{paragraph to talk about cyberwater}

%%%%%%%%%%%%%%%%%%%%%%%%% Introduce problem here %%%%%%%%%%%%%%%%%%%%%%%%%

Multiscale and multiphysics problems often need to couple different models to address their complex, interactive, and mutually influential natures. 
One classical scientific computing problem is Earth System Models, such as E3SM~\cite{e3sm-model}, which rely on highly specialized couplers to facilitate the exchange of data between participant models.
Although domain-specific couplers, like E3SM's CPL7, enable highly efficient data transfer, they are largely inflexible and cannot be utilized with other models outside of their ecosystem. 

To help domain scientists integrate various users' models, the NSF's Cyberwater framework \cite{cyberwater} is built to use a data-exchange service to connect distinct models that execute in distributed computing systems.
As a part of the framework, the Data Exchange Service (DES) is a web service-based coupler that allows the exchange of data between scientific models through a generic service component.
The Data Exchange Service promotes interoperability among models, with the exception that the models must adhere to sharing certain units of measurement.
% Transition sentence

Developers are able to build such a coupling service, however, it is a challenging task to verify the service's reliability and functionality.
Considering the criticality of the correctness of scientific computing simulations, rigorous testing methods are essential to ensure data is handled correctly and to certify the reliability of the DES Controller, as coupling logic resides in this component.

This paper is organized as follows. 
\autoref{sec:related} provides related work. 
\autoref{sec:methodology} describes our testing methodology. 
\autoref{sec:application} and \autoref{sec:results} show how we apply rigorous specification and testing to the DES Controller, with results in \autoref{sec:results}. 
Conclusion and future work are given in \autoref{sec:conc}.

%%%%%%%%%%%%%%%%%%%%%%%%%%%% Related Work %%%%%%%%%%%%%%%%%%%%%%%%%%%%
\section{Related Work}
\label{sec:related}

% \subsection{Model-Based Testing}
Finite State Machines (FSMs) are commonly used to generate test cases for event-driven software. 
In \cite{Wang_Sampath_Lei_Kacker_2008}, the authors use a sequence-based approach to test interactions of shared objects and pages in websites.
Others use FSMs to discover navigation errors in web pages \cite{Hallé_Ettema_Bunch_Bultan_2010}.
\cite{Bombarda_Gargantini_2020} uses combinatorial test generation to create initial test sequences from an FSM and repairs or discards invalid sequences.
The authors of \cite{Srivastava_Jose_Barade_Ghosh_2010} use ant colony optimization to minimize the cost of test sequences of a Markov-chain usage model.

% \subsection{Sequence-Based and specification-based Testing}
T-way sequences \cite{weighted_combin_test_tls, Kuhn_Raunak_Kacker_2023} are used in combinatorial testing to create a reliable test environment.
In \cite{Offutt_Liu_Abdurazik_Ammann_2003} the authors apply specification-based testing to cruise control software and record coverage of system interactions and state transitions.
Cayley graphs may be used, with respect to a metric, to generate full coverage test sequences as seen in \cite{Hallé_Khoury_2021}.

% \cite{Bennett_2021} builds an API testing program which builds context between chains of API endpoints and databases to prevent flaky tests.

%%%%%%%%%% Rigorous Specification and Testing Methodologies %%%%%%%%%%
\section{Rigorous Specification and Testing Methodologies}
\label{sec:methodology}

\textit{Statistical usage-based testing}, combined with \textit{sequence-based specification}, provides a rigorous testing method to systematically examine the behavior of software in all possible real-world usage scenarios, and to assess its reliability based on the testing experience.
As outlined in \cite{Poore11, Prowell99}, the benefits of statistical testing lie in weighted testing towards the most frequent operational uses of the software.
Sequence-based specification, as a \textit{black box} specification method, considers only the external inputs and outputs of the module being tested \cite{Lin10, Prowell_Poore_2003, Prowell99}.

\subsection{Sequence-Based Specification}

First, a \textit{system boundary} is defined to identify the inputs and outputs between the system, the module being tested, and the software's \textit{environment}.
The software's environment consists of the interfaces used to communicate with the system.

Next, a functional mapping is created to associate all possible input, or \textit{stimulus}, sequences with their expected outputs, or \textit{responses}, and equivalencies, if applicable, to length-lexicographically smaller sequences.
This mapping is discovered through a systematic process called \textit{sequence enumeration} and defines the test oracle for subsequent usage-based statistical testing.

To enumerate, start with the empty sequence $\lambda$ with length $0$.
Then we extend each length $n$ sequence by every possible stimulus to get all length $n+1$ sequences and consider them in lexicographical order.

For each new sequence, a decision is made to map to an expected response according to the requirements, and \textit{reduce} to a prior sequence if it takes the system to a previously seen state.
Otherwise, the sequence is designated as \textit{unreduced}, serving as an unseen state.
Some sequences are \textit{illegal} per software specifications and are denoted as $\omega$ in their response and not extended further.
Sequences which are reduced or illegal are not extended further.
Enumeration is terminated when all enumerated sequences of a certain length are reduced or illegal.

The unreduced and legal sequences are \textit{canonical sequences} which represent unique states within the system.
Canonical sequences enable us to construct a \textit{Mealy machine}, a finite-state machine composed of canonical sequences as nodes and arcs defined by stimuli and responses.

\subsection{Statistical Testing}
% \note{We create a markov-chain usage model based on the mealy machine \\
% Then we add probabilities to the arcs \\ 
% Then we have a usage model \\ 
% Then we generate test sequences from the usage model based on sampling options \\
% Sampling can be weighted, random, or coverage based with options for a minimum number of steps. 
% test cases are executed automatically from the source to the sink with each transition being a test case. \\ 
% \\ 
% we need to record what steps fail and if they're a stop failure or continue. 
% Model analysis (are we happy with the model?) -> test -> evaluate -> record -> statistical analysis
% }

% The test oracle is defined by the Mealy machine.
% For a test to pass, two conditions must be met for each stimulus in the sequence:
% 1.) the next input must be a legal stimulus for the current canonical sequence
% 2.) the output of the stimulus must be the expected output as defined in the mapping.
% A test may continue, as a fail, if the second condition is met, but the first is analogous to the software crashing.

A \textit{Markov-chain usage model} is necessary for statistical usage-based testing.
% It can be created from the Mealy machine by simply applying probabilities to each arc.
% The most common stimuli should have higher probabilities defined, allowing these functions to be tested more often.
By defining probabilities for each arc of the Mealy machine, obtained through specification, a usage model is derived.
From each usage state, higher probabilities should be defined on the most common stimuli, allowing these functions to be tested more often.
The purpose of the usage model is to characterize a population of all possible and \textit{the most frequent use cases} of the system.
To validate the model, standard Markov analysis is performed to determine if the model reflects the expected usage. % of the system under test

Test cases are then generated from the usage model by random, weighted, or coverage sampling.
A test case is a sequence of stimuli following the arcs of the usage model, starting from the \textit{source} and ending at the \textit{sink}.
At each step, one checks whether the output is expected and if the system is in the correct state.
If either output or state is incorrect, the test step fails.
If the state is incorrect and the next stimulus is illegal, then the test ends with a \textit{stop failure}, otherwise the failure step is a \textit{continue failure}.
To certify the system, quantitative measures, like \textit{Single Use Reliability}, are calculated taking into consideration the usage model, the sample of generated test cases, and test results.

\section{Applying Rigorous Specification and Testing to the Data Exchange Service}
\label{sec:application}
    The DES follows the \textit{Model-View-Controller} pattern, wherein sessions function as the model, the Data Exchange Controller serves as the controller, and clients represent the view. 
    We apply our testing methods to the controller.
    Correct functionality of the server largely depends on a session's interaction with the controller.
    To establish a clear system boundary for testing, we first derive a comprehensive set of high-level requirements from available documents and the initial system.
    % \todo{Look back at this last sentence after editing \autoref{sec:results}}.
    
\subsection{Requirements}
\vspace{-0.13in}
    The DES serves as a way for different scientific models to communicate with each other via sessions.
    Sessions are designed to be model-agnostic through user-defined parameters. 
    Information about participant model IDs, initiator and invitee IDs, variables, and variable size are required when starting a session.

    An initiator initiates a session, and invitees join sessions. 
    However, a session cannot be joined unless the invitee possesses the ID specified by the initiator.
    Variables are denoted by a user-defined integer ID serving as a key in the session's \textit{Flag Status} table telling users if data is ready to be received.
    A $0$ represents data is not present and a $1$ tells the users data is available.
    Data cannot be overwritten, as the controller will reject data for variables with a flag equal to $1$.
    Detailed requirements for handling the controller are shown in \autoref{tab:requirements}.

    \begin{table}[h!]
        \centering
        \begin{footnotesize}
        \begin{tabular}{|c|p{0.4\textwidth}|} \hline % {|p{2in}|l|}
            \textbf{Tag} & \textbf{Requirement} \\ \hline
            \textbf{1}  & \textbf{Session Creation:}\\ \hline 
            1a & The initiator shall send a request to the server to create a new session.\\ \hline 
            1b & On receiving the request the server shall create a new session.\\ \hline 
            1c & The server shall send a reply/acknowledgment message to the initiator.\\ \hline 
            \textbf{2} & \textbf{Joining Sessions:}\\ \hline 
            2a & The client shall send a request to the server to join a session.\\ \hline 
            2b & The server shall check the request to see if the session exists.\\ \hline 
            2c & The server shall check the request to see if the client is an invitee in the existing session.\\ \hline 
            \textbf{3} & \textbf{Sending data:}\\ \hline 
            3a & The client shall be in a session in order to send data.\\ \hline 
            3b & The client shall send a send-data request to the server.\\ \hline 
            3c & The server shall reject send-data requests for sessions that don’t exist.\\ \hline
            \textbf{4} & \textbf{Receiving data:}\\ \hline 
            4a & The client shall be in a session in order to receive data.\\ \hline 
            4b & The client shall send a receive-data request to the server.\\ \hline
            4c & The server shall check (via the data’s flag) if data is present.\\ \hline 
            \textbf{5} & \textbf{Ending Sessions:}\\ \hline 
            5a & Either client shall send an end-session request if data exchange is no longer needed from the client.\\ \hline 
            5b & The server shall reject end-session requests for sessions that do not exist.\\ \hline
            5c & The server shall reject end-session requests from clients not in the session.\\ \hline 
            \textbf{6} & \textbf{Session Requirements:}\\ \hline 
            6a & The session shall be created first.\\ \hline
            6b  & A session shall be independent of each other. \\ \hline
            6c  & A new session shall have default flags set to 0. \\ \hline
            % 6f  & The session shall have a finite positive number of data flags. Enumerating all the data flags is not productive. This enumeration will only keep track if some data flag is changed.  \\ \hline
        \end{tabular}
        \end{footnotesize}
        \caption{Excerpts from Data Exchange Controller requirements}
        \label{tab:requirements}
    \end{table}

\vspace{-0.27in}
\subsection{System Boundary}

\begin{figure}[h]
    \centering
    \includegraphics[width=0.45\textwidth]{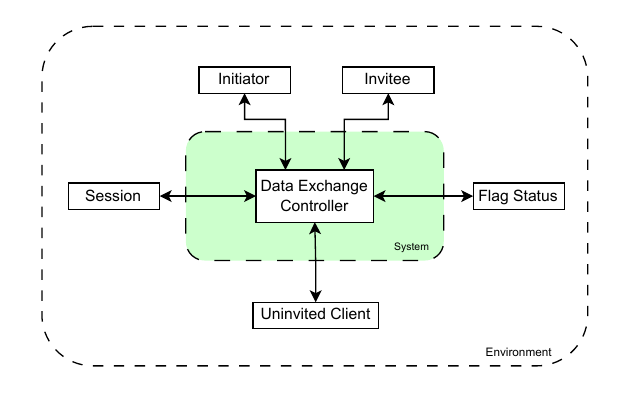}
    \caption{The system boundary for the Data Exchange Controller and its interactions with the environment}
    \label{fig:software_env}
    \end{figure}

    The system boundary is defined around the Data Exchange Controller, as all inputs and outputs route through it and it contains most of the logic within the package.
    The system boundary and software's environment are shown in \autoref{fig:software_env}.
    The uninvited client is included due to requirements of joining sessions.
    While sessions do not include the uninvited client, one can attempt to join a session.
      
\subsection{Enumeration and Mealy Machine}

    % The functional mapping represents the HTTP endpoints that manipulate the session states; HTTP GET endpoints are not included.
    Stimuli and responses are identified and grouped by HTTP POST endpoints, which manipulate session states.
    \autoref{tab:inputs} shows the mapping between stimulus keys, HTTP POST endpoints, and responses on error or success.

    \begin{table*}[h]
    \centering
    \begin{tabular}{clll}
    \textbf{Stimulus Key}  & \textbf{Endpoint} & \textbf{Success Response} & \textbf{Error Response} \\
        $C$ & /create\_session & create\_session, create\_session\_ack & create\_session\_err \\
        $E$ & /end\_session    & end\_session\_ack, clear\_flag\_and\_data & end\_session\_err \\ 
        $J$ & /join\_session   &  join\_session, join\_session\_ack & join\_session\_err \\
        $R$ & /receive\_data   & retrieve\_data, recv\_data\_ack, update\_flag(0) & recv\_data\_err \\
        $S$ & /send\_data     & send\_data\_ack, store\_data, update\_flag(1) & send\_data\_err \\
    \end{tabular}
    \caption{Endpoints and their corresponding stimulus keys and responses}
    \label{tab:inputs}
    \end{table*}

    % \begin{table*}[!ht]
    % \centering
    % \begin{tabular}{llll}
    % \textbf{Stimulus Key}  & \textbf{Endpoint} & \textbf{Success Response} & \textbf{Error Response} \\
    %     $C$ & /create\_session & $c_e$, $c_a$ & $c_e$ \\
    %     $E$ & /end\_session    & $e_a$, clr & $e_e$ \\ 
    %     $J$ & /join\_session   & $j_s$, j_a & $j_e$ \\
    %     $R$ & /receive\_data   & $r_d$, $r_a$, $uf(0)$ & $r_e$ \\
    %     $S$ & /send\_data     & $s_a$, $str$, $uf(1)$ & $s_e$ \\
    % \end{tabular}
    % \caption{Endpoints and their corresponding stimulus keys and responses}
    % \label{tab:inputs}
    % \end{table*}

    To further refine the enumeration, \textit{predicates} are used to specify what a stimulus should do.
    We specify \textit{invalid} inputs with an $_f$ and \textit{valid} inputs with a $_t$, e.g., $R_f$ and $R_t$ are invalid and valid receive data requests respectively.
    Error responses are used to check invalid inputs, such as an uninvited client attempting to join a session. 
    Predicate refinements may not be needed for some sequences during enumeration. 
    For instance, any $/receive\_data$ request is invalid if data is not present.
    
    \begin{table*}[h]
        \centering
        \begin{footnotesize}
        \begin{tabular}{|l|l|l|l|} \hline 
        \textbf{Sequence}  & \textbf{Response} & \textbf{Equivalence} & \textbf{Trace} \\ \hline 
            \rowcolor{skyblue}$\lambda$       & 0& &          Method\\ \hline 
            $C_f$  & create\_session\_err &  $\lambda$   & 1c, 1g \\ \hline  
            \rowcolor{skyblue}$C_t$  & create\_session, create\_session\_ack &     & 1b, 1c, 1d, 1e, 1f, 1g \\ \hline  
            $E$  & $\omega$ & &      6a\\ \hline  
            $J$  & $\omega$ & &      6a\\ \hline  
            $R$  & $\omega$ & &      6a\\ \hline  
            $S$  & $\omega$ & &      6a\\ \hline  
            $C_tC  $ &  $\omega$ & &      6b\\ \hline  
            $C_tE  $ &  end\_session\_ack, clear\_flag\_and\_data& $\lambda$& 5d, 5f\\ \hline  
            $C_tJ_f$ &  join\_session\_err& $C_t$& 2b, 2c, 2d, 2e, 5e\\ \hline  
            \rowcolor{skyblue}$C_tJ_t$ &  join\_session\_ack& & 2f\\ \hline  
            $C_tR  $ &  recv\_data\_err& $C_t$& 4a, 4c, 4d, 4e\\ \hline 
            $C_tS_f  $ &  send\_data\_err& $C_t$ & 3a, 3c, 3d, 3e, 6f\\ \hline  
            \rowcolor{skyblue}$C_tS_t  $ &  send\_data\_ack, store\_data, update\_flag(1)& & 3f, 3g, 3h, 6c\\ \hline  
            $C_tJ_tC$         & $\omega$ & & 6b\\ \hline  
            \rowcolor{skyblue}$C_tJ_tE$         & end\_session\_ack& & 5d\\ \hline  
            $C_tJ_tJ$         & $\omega$ & & 6d\\ \hline 
            $C_tJ_tR$         & recv\_data\_err& $C_tJ_t$& 4a, 4c, 4d, 4e\\ \hline  
            $C_tJ_tS_f  $ &  send\_data\_err& $C_tJ_t$ & 3a, 3c, 3d, 3e, 6f\\ \hline  
            \rowcolor{skyblue}$C_tJ_tS_t$         & send\_data\_ack, store\_data, update\_flag(1)& & 3f, 3g, 3h, 6c\\ \hline 
            % $C_tS_tC$  &  $\omega$ & & 6b\\ \hline  
            % $C_tS_tE$  &  end\_session\_ack, clear\_flag\_and\_data& $\lambda$ & 5d, 5f\\ \hline  
            % $C_tS_tJ_f$         & join\_session\_err& $C_tS_t$& 2b, 2c, 2d, 2e, 5e\\ \hline  
            % $C_tS_tJ_t$         & join\_session\_ack& $C_tJ_tS_t$& 2f\\ \hline  
            % $C_tS_tR_f$         & recv\_data\_err& $C_tS_t$& 4a, 4c, 4d, 4e  \\ \hline  
            % $C_tS_tR_t$         & recv\_data\_ack, retrieve\_data, update\_flag(0)& $C_t$& 4f, 4g\\ \hline  
            % $C_tS_tS_f$         & send\_data\_err& $C_tS_t$&  3a, 3c, 3d, 3e, 6f\\ \hline 
            % $C_tS_tS_t$         & send\_data\_ack, store\_data, update\_flag(1)   & $C_tS_t$& 3f, 3g, 3h, 6c, 6e, 6f\\ \hline  
            % $C_tJ_tEC  $      & $\omega$ &        & 6b\\ \hline    
            % $C_tJ_tEE  $      & end\_session\_ack, clear\_flag\_and\_data        & & 5d, 5f\\ \hline   
            % $C_tJ_tEJ  $      & join\_session\_err &  $C_tJ_tE$& 2e, 5e\\ \hline   
            % $C_tJ_tER  $      & recv\_data\_err& $C_tJ_tE$& 4a, 4c, 4d\\ \hline   
            \multicolumn{4}{|c|}{$\cdots$} \\ \hline
            \rowcolor{skyblue} $C_tJ_tES_t  $      & send\_data\_ack, store\_data, update\_flag(1)& & 3f, 3g, 3h, 6c, 6e, 6g\\ \hline   
    %         $CS_tC$  &  $\omega$ & & 6b\\ \hline  
    %         $CS_tE$  &  end\_session\_ack, clear\_flag\_and\_data& $\lambda$ & 5d, 5f\\ \hline  
    %         $CS_tJ_f$         & join\_session\_err& $CS_t$& 2b, 2c, 2d, 2e, 5e\\ \hline  
    %         $CS_tJ_t$         & join\_session\_ack& $CJ_tS_t$& 2f\\ \hline  
    %         $CS_tR_f$         & recv\_data\_err& $CS_t$& 4a, 4c, 4d, 4e  \\ \hline  
    %         $CS_tR_t$         & recv\_data\_ack, retrieve\_data, update\_flag(0)& $C$& 4f, 4g\\ \hline  
    %         $CS_tS_f$         & send\_data\_err& $CS_t$&  3a, 3c, 3d, 3e, 6f\\ \hline 
    %         $CS_tS_t$         & send\_data\_ack, store\_data, update\_flag(1)   & $CS_t$& 3f, 3g, 3h, 6c, 6e, 6f\\ \hline  
    %         $CJ_tEC  $      & $\omega$ &        & 6b\\ \hline    
    %         $CJ_tEE  $      & end\_session\_ack, clear\_flag\_and\_data        & & 5d, 5f\\ \hline   
    %         $CJ_tEJ  $      & join\_session\_err &  $CJ_tE$ & 2e, 5e\\ \hline   
    %         $CJ_tER  $      & recv\_data\_err& $CJ_tE$ & 4a, 4c, 4d\\ \hline   
            \multicolumn{4}{|c|}{$\cdots$} \\ \hline
            $C_tJ_tES_tS_f$     & send\_data\_err& $C_tJ_tES_t$& 3a, 3c, 3d, 3e, 6f\\ \hline
            $C_tJ_tES_tS_t$     & send\_data\_ack, store\_data, update\_flag(1)& $C_tJ_tES_t$& 3f, 3g, 3h, 6c, 6e, 6f\\ \hline 
        \end{tabular}
        \end{footnotesize}
        \caption{Excerpts from the enumeration table. Canonical sequences are shown in colored rows.}
        \label{tab:enum_table}
    \end{table*}

    We enumerate sequences of stimuli in length-lexicographical order that represent histories of events received by the controller. 
    An excerpt of the enumeration table is presented in \autoref{tab:enum_table} with canonical sequences highlighted in blue. 
    The first sequence  ($\lambda$) is canonical, representing the lack of inputs, or more specifically, no session.
    The next sequence is $C_f$.
    $C_f$ is a create session request with bad input data, so an error response is returned and the sequence is logically equivalent to $\lambda$ as no new session is created.
    The next sequence ($C_t$) is canonical, which returns new session information in the acknowledgment. 
    The remaining length one sequences are all illegal since a session is required to perform their actions. 
    Because of this, $\omega$ is shown in response denoting these sequences are illegal. 
    
    Enumeration continues with a length of two starting at $C_tC$, as $C_t$ is the only canonical sequence of length one that is extended further. 
    A predicate is not specified for the second $C$ as creation of a new session is not possible within a session. 
    % a new session would be independent of the first or an invalid request would not create a new one; either case does not change the state of the first stimulus.
    The next sequence is $C_tE$; creating and deleting a session (only one user is connected). It is made equivalent to $\lambda$.
    There are two canonical sequences of length two: $C_tJ_t$ and $C_tS_t$ (creating session with a join or send request).
    
    The new canonical sequences are extended, like $C_t$, with each stimulus.
    The extended sequences are considered in lexicographical order for response mappings and equivalence decisions. 
    The enumeration process terminates when there are no new canonical sequences to extend, as seen with length 5 sequences ending with $C_tJ_tES_tS_t$.
    % The process described with $C_t$ is repeated for the two new canonical sequences and terminates with the final row $C_tJ_tES_tS_t$ as no new irreducible sequences appear.
    
\begin{figure}[h]
    \centering
    \includegraphics[width=0.5\textwidth]{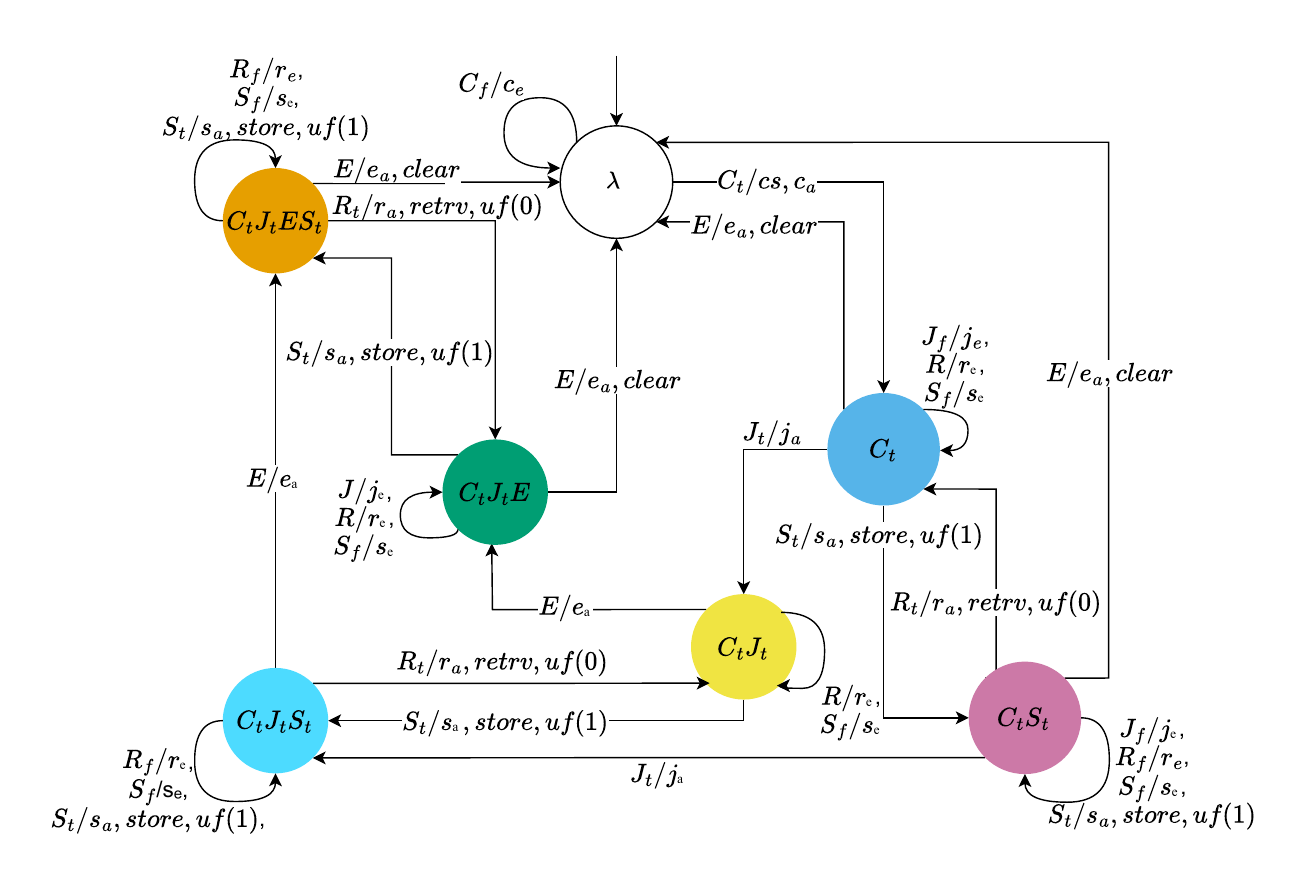}
    \caption{Mealy Machine for sessions in the Data Exchange Service. Shorthand notation is used for stimuli and responses in the form $Stim./Resp.$. }
    \label{fig:mealy}
    \end{figure}

    With the completed enumeration, a Mealy machine is created using canonical sequences as states. 
    Rows of the enumeration table define transitions among states and outputs on the arcs.
    The Mealy machine for the Data Exchange Controller, in \autoref{fig:mealy}, is used to generate test cases and define the test oracle. 
    The Mealy machine is derived manually from the enumeration following a defined procedure \cite{Lin10, Prowell99}. 
    %determine if the controller has put a given session in the proper state during testing.
    A Markov-chain usage model is created by assigning probabilities to every arc from every state.%, e.g., it can be assumed $S$ is a more common input for state $C$ than $E$, so a weight of $0.8$ and $0.2$ can be defined respectively. 

    \subsection{Canonical Sequence Analysis and Test Oracle}
    The completed sequence-based specification defines the controller's test oracle.
    During testing, both the session's states and controller's outputs are verified.
    Testing responses is straightforward as HTTP codes or returned output may be tested similarly to unit test. To test the system's internal states, canonical sequences are used.
    By \textit{Canonical Sequence Analysis} each canonical sequence can be determined using session attributes.
    Intuitively, each attribute (2nd - 5th column headers) in \autoref{tab:can_seq_analysis} represents a state variable captured by the canonical sequences.
    Attributes are read using HTTP GET requests to view flag table or session occupancy values. 
    % Implicitly, GET requests are tested with POST requests. 
    
    \begin{table}[h]
        \centering
        \begin{footnotesize}
        \begin{tabular}{|l|c|c|c|c|} \hline 
             \textbf{Canonical Seq.} & \textbf{Created} & \textbf{Joined} & \textbf{Data Sent} & \textbf{Partial End}\\ \hline 
             $\lambda$&  0&  -&  -& -\\ \hline 
             $C_t$&  1&  0&  0& -\\ \hline 
             $C_tJ_t$&  1&  1&  0& 0\\ \hline 
             $C_tS_t$&  1&  0&  1& -\\ \hline 
             $C_tJ_tE$&  1&  1&  0& 1\\ \hline 
             $C_tJ_tS_t$&  1&  1&  1& 0\\ \hline 
             $C_tJ_tES_t$&  1&  1&  1& 1\\ \hline
        \end{tabular}
        \caption{The Canonical Sequence Analysis table represents a list of features that describe the state of the session. `-' means the feature does not apply to the state.}
        \label{tab:can_seq_analysis}
        \end{footnotesize}
    \end{table}

    % All but 'Data Sent' is checked using the session's status attribute.

\section{Automated Statistical Testing of the Data Exchange Controller}
\label{sec:results}

    To test the Data Exchange Controller, we implement an automated test workflow using a variety of tools, including the J Usage Model Builder Library (JUMBL) \cite{Prowell_2003}, with the Markov-chain usage model written in The Model Language (TML) \cite{prowellTML}, and a Python program developed to interface with the JUMBL.

\subsection{Automated Testing Program}

    % \label{lst:tml}

    % \begin{figure}
    %     \centering
    %     \includegraphics[width=0.45\textwidth]{figs/TML.pdf}
    %     \caption{Excerpt from Data Exchange TML File}
    %     \label{fig:tml}
    % \end{figure}

\vspace{-0.1in}
    % The Markov-chain usage model is written in \textit{The Model Language} (TML) \cite{prowellTML}.
    TML is a language for describing Markov-chain usage models. The controller's TML file is used as an input to the JUMBL (our statistical testing tool) for generating test cases.
    An excerpt of the controller's TML is shown here:
\lstset{captionpos=b, basicstyle=\ttfamily\footnotesize}
\begin{lstlisting}
 1. ($ fill (1) $)
 2. model DataExchangeAPI
 3.    
 4. source [lambda]
 5.   ($0.01$)  "C_f/c_e"              [lambda]
 6.             "C_t/cs, c_a"          [C_t]
 7.  
 8. [C_t]
 9.  ($0.1$)   "E/e_a, clear"          [Exit]
10.  ($0.005$) "J_f/j_e"               [C_t]
11.            "J_t/j_a"               [C_tJ_t]
12.  ($0.01$)  "R_f/r_e"               [C_t]
13.  ($0.01$)  "R_t/r_e"               [C_t]
14.  ($0.01$)  "S_f/s_e"               [C_t]
15.            "S_t/s_a, store, uf(1)" [C_tS_t]
\end{lstlisting}
    with \texttt{[lambda]} and \texttt{[C\_t]} on lines 4 and 8 representing two states, and the lines below (5-6, and 9-15) are probabilities, stimuli, responses, and state transitions respectively.
    % To handle sequence generation and model and test analyses, we use the J Usage Model Builder Library (JUMBL)\cite{Prowell_2003}.

    To handle test sequence generation and model and test analyses, we use the JUMBL tool.
    Model analysis computes statistics of the usage model following standard Markov analysis. 
    These statistics have interpretation in software testing which can be used to validate the model. 
    JUMBL finds session state occurrence and occupancy, showing how often a state is visited or a stimulus is encountered in long-run random testing.
    
    Our tests are automatically generated with three different sampling options: weighted, random, and minimum coverage. 
    Weighted sampling picks the most probable paths in the usage model using the product of arc probabilities.
    Random sampling uses the probabilities defined on each arc to generate the next stimulus.
    Minimum coverage creates a set of sequences with minimum total steps to cover each node and each arc of the usage model.

    The main automated testing functionality lies outside of the JUMBL tool. 
    Unlike in previous work \cite{Lin2015}, where we annotate the usage model with testing scripts and then generate executable test cases from the model, 
    we write a Python script to run the software environment, test oracle, and functions interfacing with the JUMBL to automate test case generation, execution, evaluation, and recording of test results.
    
    Sequences are exported from the JUMBL test record and parsed by our Python script,
    % Functions are mapped to each stimulus and responses are mapped to an expected output.
    which generates test inputs given by the sequence and verifies outputs associated with each stimulus' response as well as the session states.
    Pass and fail information, including stop failures, is recorded for every step of the executed test sequence.
    % After parsing and constructing test cases, all tests are run automatically.
    % The oracle and compares the actual output with the expected output and determines the state of of the session by checking against \autoref{tab:can_seq_analysis}.
    % Because sessions are independent, testing is trivially parallelizable between sessions, so test suites are run quickly. 
    
    When testing is completed, JUMBL runs statistical analysis on the test results.
    Test case analysis computes statistics of reliability estimates, like Single Use Reliability, and information theoretic measures, like Relative Kullback Discrimination, to assist the test stopping criteria.  {\it Single Use Reliability} is defined as the probability of a randomly selected use being successful. {\it Relative Kullback Discrimination} reflects if testing approximates the expected use as described by the usage model. These two are among the most important statistics to consider regarding management decisions.
    
\subsection{Results}
    \vspace{-8.5pt}
    %To test the validity of our method, the test script is run during the development of the DES. 
    We apply automated statistical testing to the iterative development of the DES.
    While following test-driven development, pre-developed unit tests missed bugs resulting from some specific input sequences. 
    The automated test script helps expose bugs not found by unit tests. 
    New unit tests are written to address the conditions found by statistical testing and further analysis is performed to discover more erroneous usage scenarios.

    Tests are run on three versions of the Data Exchange Controller. 
    The old version is a prototype to demonstrate the idea of a flexible model-coupling server and was written without formal requirements.
    The new version is written with requirements derived from the old version, and a newer, fixed version is included with bugs identified and fixed during statistical testing.

    % \begin{itemize}
    %     \item $C_tJ_tE_{1st}J_t$ Joining a partially ended session made it active again. /join\_session was missing a condition which checked the session status.
    %     \item Data was received even when the flag was 0. /receive\_data didn't check for flags, and only checked if data was null.
    %     \item Creating a session was possible with mismatching lengths of sizes and variables or empty lists. /create\_session didn't check for lengths in lists.
    % \end{itemize}
    The bugs found during statistical testing of the new version are shown in \autoref{tab:bugs}, with fragments of the failed sequences shown in the first column.
    Reasons of failure were identified with human inspection.
        
    \begin{table*}[t]
        \centering
        \begin{tabular}{|p{1.3in}|p{1.6in}|p{1.6in}|p{1.6in}|}
        \hline 
         \textbf{Failed Test Fragment} & \textbf{Reason of Failure} & \textbf{Identified Bugs} & \textbf{Implemented Fixes} \\ \hline 
           $C_tJ_tEJ_t$  & User was able to join after an end session request when an error should be returned. & The server join function did not check session status to see if a user left.  & Server join function was updated to check session status for partial-end. \\ \hline 
           $C_tS_tR_tR_t$ & Second receive request returned data when an error should be returned. & Server receive function only checked if data existed, but did not check relevant flag. & Server receive function was updated to check if relevant flag is $1$.\\ \hline 
           $C_f$  & The function associated with $C_f$ is missing input validation. Session was created with missing information. & Function did not check for missing information. & Input validation added to the function to check request for valid parameters. \\ \hline 
        \end{tabular}
        \caption{Test fragments caused errors on the new version. The final stimulus failed during testing.}
        \label{tab:bugs}
    \end{table*}

    Our generated test suite includes four min-coverage tests, $200$ weighted tests, and $5,000$ random tests.
    The reliability of the DES is certified by testing until a threshold of 99\% Single Use Reliability is achieved.
    % We test an older version of the DES using the same $3800$ tests, but with a slightly modified testing script for compatibility. 
    With relatively few stimuli and many tests, some sequences may be repeats or contain repeated components.
    This is normal for statistical testing because a random sample can contain many tests which are not necessarily unique.
    In a real-world scenario, the expected use cases are repeated frequently.

    \begin{footnotesize}
    \begin{table*}[h!]
        \centering
        \begin{tabular}{|lr|rr|rr|rr|} \hline
        \multicolumn{2}{|c|}{\textbf{Total Generated}} & \multicolumn{2}{|c|}{\textbf{Old}} & \multicolumn{2}{|c|}{\textbf{New}} & \multicolumn{2}{|c|}{\textbf{Fixed}} \\ \hline 
        \textbf{Stimulus/Response}  & \textbf{Generated} &  \textbf{Executed} & \textbf{Failed}  & \textbf{Executed} & \textbf{Failed} & \textbf{Executed} & \textbf{Failed} \\ \hline
        $C_f/c_e$            & $56$     & $56$     & $56$   & $56$     & $56$  & $56$     & $0$ \\ \hline 
        $C_t/c_a,c_s$        & $5,204$  & $5,148$  & $0$    & $5,148$  & $0$   & $5,204$  & $0$ \\ \hline
        $E/e_a$              & $4,194$  & $4,097$  & $2,351$& $4,145$  & $0$   & $4,194$  & $0$ \\ \hline      
        $E/e_a,clear$        & $5,204$  & $5,094$  & $70$   & $5,143$  & $77$  & $5,204$  & $0$ \\ \hline    
        $J_f/j_e$            & $125$    & $120$    & $120$  & $121$    & $0$   & $125$    & $0$ \\ \hline            
        $J_t/j_a$            & $4,194$  & $4,097$  & $0$    & $4,145$  & $0$   & $4,194$  & $0$ \\ \hline           
        $J_t/j_e$            & $86$     & $78$     & $78$   & $82$     & $82$  & $86$     & $0$ \\ \hline           
        $R_f/r_e $           & $465$    & $451$    & $77$   & $456$    & $3$   & $465$    & $0$ \\ \hline           
        $R_t/r_a,retrv,uf(0)$& $11,143$ & $10,867$ & $1,531$& $11,005$ & $79$  & $11,143$ & $0$ \\ \hline            
        $R_t/r_e$            & $502$    & $490$    & $230$  & $493$    & $376$ & $502$    & $0$ \\ \hline         
        $S_f/s_e $           & $434$    & $424$    & $424$  & $427$    & $4$   & $434$    & $0$ \\ \hline    
        $S_t/s_a,store,uf(1)$& $18,271$ & $17,865$ & $4,669$& $18,060$ & $119$ & $18,271$ & $0$ \\ \hline 
        \textbf{Total Stimuli/Responses} & $49,878$ & $48,787$ & $9,606$ & $49,281$ & $796$ & $49,878$  & $0$ \\ \hline
        \textbf{Total Tests} & $5,204$ &  $5,204$ & $3,786$& $5,204$  & $482$ & $5,204$  & $0$    \\ \hline \hline  
        \multicolumn{2}{|l|}{\textbf{Single Use Reliability}} & \multicolumn{2}{|c|}{$0.276368459$} & \multicolumn{2}{|c|}{$0.863924505$} & \multicolumn{2}{|c|}{$0.992865083$}  \\ \hline
        \multicolumn{2}{|l|}{\textbf{Relative Kullback Discrimination}} & \multicolumn{2}{|c|}{$35.4876156E-3\%$} & \multicolumn{2}{|c|}{$37.2213956E-3\%$} & \multicolumn{2}{|c|}{$36.333333E-3\%$} \\ \hline
        \end{tabular}
        \caption{Test results for the three versions of the Data Exchange Controller}
        \label{tab:results}
    \end{table*}
    \end{footnotesize}
    Testing results, shown in \autoref{tab:results}, include Single Use Reliabilities, Relative Kullback Discriminants, numbers of stimuli and tests generated/executed/failed.
    The test analysis report includes node and arc statistics and reliabilities, but they are not included here for brevity. 
    The mismatch between generated and executed stimuli is due to stop failures, where failed stimuli did not modify the session state correctly because of an error.
    
    The fixed version passes every test. The new version passes $90.7\%$, and the old version passes only $27.2\%$ due to bugs and also requirements changes introduced from the old version to the new version.
    Using rigorous specification and testing, the Single Use Reliability is improved by $13\%$ over the new version of the controller, and $72\%$ over the old version.
    The Relative Kullback Discrimination remains low for all three versions, indicating the testing experience is approximating the expected uses.
    We certify the reliability of the Data Exchange Controller after showing it passing $49,878$ valid and invalid inputs and achieving $99.3\%$ Single Use Reliability.

\section{Conclusion and Future Work}
\label{sec:conc}
\vspace{-0.13in}
In this paper, we certify the reliability of the Data Exchange Controller using sequence-based specification and usage-based statistical testing.
Different versions of the Data Exchange Controller are compared to illustrate the effectiveness of our approach for testing stateful HTTP sessions.
The newer, fixed version is shown to be reliable and robust while coupling scientific models.

%%%%%%%%% Future Work %%%%%%%%%
This work uses specialized code to interface with the JUMBL tool.
In the future, a generalized framework for fully-automated test case generation, execution, and evaluation can be developed to expedite the implementation of statistical testing regardless of the application.
The rigorous software engineering methodologies we use are also applicable to testing other scientific software infrastructure.

\vspace{-0.13in}
\section*{Acknowledgments}
\vspace{-0.13in}
The authors thank Ayush Lodha and Dr. Yao Liang at Indiana University Indianapolis for an earlier version of the DES software to test.
This work is generously funded by NSF under Grant \#2209834 and Grant \#2209835.

\bibliographystyle{latex8}
\bibliography{refs}

\begin{thebibliography}{10}\setlength{\itemsep}{-1ex}\small

\bibitem{Bombarda_Gargantini_2020}
A.~Bombarda and A.~Gargantini.
\newblock An automata-based generation method for combinatorial sequence testing of finite state machines.
\newblock In {\em 2020 IEEE International Conference on Software Testing, Verification and Validation Workshops (ICSTW)}, page 157–166, Oct. 2020.

\bibitem{cyberwater}
R.~Chen, F.~Li, D.~Bieger, F.~Song, Y.~Liang, D.~Luna, R.~Young, X.~Liang, and S.~Pamidighantam.
\newblock Cyberwater: An open framework for data and model integration in water science and engineering.
\newblock In {\em Proceedings of the 31st ACM International Conference on Information \& Knowledge Management}, CIKM'22, page 4833–4837. ACM, 2022.

\bibitem{e3sm-model}
{E3SM Project}.
\newblock {Energy Exascale Earth System Model (E3SM)}.
\newblock [Computer Software] \url{https://dx.doi.org/10.11578/E3SM/dc.20240301.3}, May 2025.

\bibitem{weighted_combin_test_tls}
B.~Garn, D.~E. Simos, F.~Duan, Y.~Lei, J.~Bozic, and F.~Wotawa.
\newblock Weighted combinatorial sequence testing for the {TLS} protocol.
\newblock In {\em 2019 IEEE International Conference on Software Testing, Verification and Validation Workshops (ICSTW)}, pages 46--51, 2019.

\bibitem{Hallé_Ettema_Bunch_Bultan_2010}
S.~Hallé, T.~Ettema, C.~Bunch, and T.~Bultan.
\newblock Eliminating navigation errors in web applications via model checking and runtime enforcement of navigation state machines.
\newblock In {\em Proceedings of the IEEE/ACM International Conference on Automated Software Engineering}, page 235–244, Antwerp Belgium, Sept. 2010. ACM.

\bibitem{Hallé_Khoury_2021}
S.~Hallé and R.~Khoury.
\newblock Test sequence generation with {C}ayley graphs.
\newblock In {\em 2021 IEEE International Conference on Software Testing, Verification and Validation Workshops (ICSTW)}, page 182–191, Apr. 2021.

\bibitem{Kuhn_Raunak_Kacker_2023}
D.~R. Kuhn, M.~S. Raunak, and R.~N. Kacker.
\newblock Ordered t-way combinations for testing state-based systems.
\newblock In {\em 2023 IEEE International Conference on Software Testing, Verification and Validation Workshops (ICSTW)}, page 246–254, Apr. 2023.

\bibitem{Lin2015}
L.~Lin, J.~He, and Y.~Xue.
\newblock An automated testing framework for statistical testing of {GUI} applications.
\newblock In {\em Proceedings of the 27th International Conference on Software Engineering and Knowledge Engineering (SEKE)}, pages 72--79, 2015.

\bibitem{Lin10}
L.~Lin, S.~J. Prowell, and J.~H. Poore.
\newblock An axiom system for sequence-based specification.
\newblock {\em Theoretical Computer Science}, 411(2):360--376, 2010.

\bibitem{Offutt_Liu_Abdurazik_Ammann_2003}
J.~Offutt, S.~Liu, A.~Abdurazik, and P.~Ammann.
\newblock Generating test data from state‐based specifications.
\newblock {\em Software Testing, Verification and Reliability}, 13(1):25–53, Jan. 2003.

\bibitem{Poore11}
J.~H. Poore, L.~Lin, R.~Eschbach, and T.~Bauer.
\newblock Automated statistical testing for embedded systems.
\newblock In J.~Zander, I.~Schieferdecker, and P.~J. Mosterman, editors, {\em Model-Based Testing for Embedded Systems in the Series on Computational Analysis and Synthesis, and Design of Dynamic Systems}. CRC Press-Taylor \& Francis, 2011.

\bibitem{prowellTML}
S.~Prowell.
\newblock {TML}: A description language for {M}arkov chain usage models.
\newblock {\em Information and Software Technology}, 42(12):835--844, 2000.

\bibitem{Prowell_2003}
S.~Prowell.
\newblock {JUMBL}: A tool for model-based statistical testing.
\newblock In {\em Proceedings of the 36th Annual Hawaii International Conference on System Sciences}, Jan. 2003.

\bibitem{Prowell_Poore_2003}
S.~Prowell and J.~Poore.
\newblock Foundations of sequence-based software specification.
\newblock {\em IEEE Transactions on Software Engineering}, 29(5):417–429, May 2003.

\bibitem{Prowell99}
S.~J. Prowell, C.~J. Trammell, R.~C. Linger, and J.~H. Poore.
\newblock {\em Cleanroom Software Engineering: Technology and Process}.
\newblock Addison-Wesley, Reading, MA, 1999.

\bibitem{Srivastava_Jose_Barade_Ghosh_2010}
P.~R. Srivastava, N.~Jose, S.~Barade, and D.~Ghosh.
\newblock Optimized test sequence generation from usage models using ant colony optimization.
\newblock {\em International Journal of Software Engineering \& Applications}, 1(2):14–28, Apr. 2010.

\bibitem{Wang_Sampath_Lei_Kacker_2008}
W.~Wang, S.~Sampath, Y.~Lei, and R.~Kacker.
\newblock An interaction-based test sequence generation approach for testing web applications.
\newblock In {\em 2008 11th IEEE High Assurance Systems Engineering Symposium}, page 209–218, Dec. 2008.

\end{thebibliography}

\end{document}